\documentclass[aps,prc,twocolumn,showpacs,amsmath,preprintnumbers,amssymb,groupedaddress,superscriptaddress]{revtex4}
\usepackage{bbm}

\usepackage{graphicx}
\usepackage{dcolumn}
\usepackage{bm}
\usepackage{multirow}
\usepackage{hhline}

\begin{document}

\preprint{}

\title{Nuclear Stopping and Sideward Flow Correlation from $0.35A$ to $200A$ GeV }

\author{Xiao-Feng~Luo}
\thanks{contact author: science@mail.ustc.edu.cn}
\author{Ming~Shao}
\author{Xin~Dong}
\author{Cheng Li}
\affiliation{University of Science and Technology of China, Hefei ,
Anhui, 230026, China}

\begin{abstract}
The correlation between the nuclear stopping and the scale invariant
nucleon sideward flow from SIS/GIS to SPS/CERN energies is studied
within Ultra-relativistic Quantum Molecular Dynamics (UrQMD). The
universal behavior of the two experimental observables for various
colliding systems and scale impact parameters are found to be highly
correlated with each other. As there is no phase transition
mechanism involved in the UrQMD, the correlation may be broken down
by the sudden change of the bulk properties of the nuclear matter,
such as the formation of Quark-gluon plasma (QGP), which can be
employed as a QGP phase transition signal in high energy heavy ion
collisions. Furthermore, we also point out that the appearance of
breaking down of the correlation may be a powerful tool to search
for the critical point on the QCD phase diagram.

\end{abstract}


\date{\today{}}

\maketitle In the recent years, the main aim of ultra-relativistic
high energy heavy ion collisions (HICs) performed at SPS/CERN
($\sqrt{{s}_\mathrm{NN}}\sim10A$ GeV) and RHIC/BNL
($\sqrt{{s}_\mathrm{NN}}\sim200A$ GeV) is to search a new form of
matter with partonic degrees of freedom, the so-called Quark-gluon
plasma (QGP) \cite{QGP1,QGP2,QGP3,QGP4}. Although great efforts have
been made, no dramatic changes of experimental observables, such as
jet-quenching, elliptic flow and strangeness enhancement, have been
observed yet and it is hard to make a solid conclusion for the
happening of the QGP phase transition \cite{No_dramatic}. Recently,
the energy scan program is proposed for RHIC/BNL to perform HICs
experiments with lower c.m. energy to search for the critical point
\cite{criRHIC1,criRHIC2,criRHIC3}, which is as a endpoint of the
first order phase transition line on the QCD phase diagram. If the
critical point exists, it should appear on the QGP transition
boundary at higher baryon chemical potential and lower colliding
energy \cite{lattic1,lattic2}. To extract the QGP phase transition
signal, a large amount of possible experimental probes, such as
particle ratio and collective flow {\it etc.} have been proposed.
The time evolution of temperature and baryon chemical potential of
the different colliding nuclei with various colliding energy would
be mapping much broader $T-\mu_{B}$ region on the QCD phase diagram
than a single nuclei. However, it is complicated to uniformly and
systematically obtain unambiguous experimental signal for QGP phase
transition and also mark the location of the critical point on the
QCD phase diagram, one of the possible choice is to keep insight in
universal correlation pattern of two experimental observables for
various colliding systems (system size and beam energy). Thus, the
complication of various colliding systems dependence of searching
the phase transition signals can be reduced.

In this work, the correlation between the nuclear stopping and scale
invariant nucleon sideward flow within the framework of UrQMD model
from SIS/GSI to SPS/CERN energies for various scale impact
parameters $0<b_{0}=b/b_{max}<1$ have been found, not just for
global fixed impact parameters as in Ref.\cite{correlation}. The
scale invariant nucleon sideward flow is defined as
$\stackrel{\sim}{F}(b_{0})={\partial{(<p_{cm}^{x}/A>/p_{cm}^{proj})}}/{\partial{(y_{cm}/y_{cm}^{proj})}}{\big
|_{[-1, 1]}} $ , first proposed in Ref. \cite{scaleflow}, the linear
fitting slope of the normalized rapidity dependence of the
normalized average in reaction plane transverse momentum with
fitting range of -1$<y_{cm}/y_{cm}^{proj}<$1, where $<p_{cm}^{x}/A>$
is the average transverse momentum projected in the reaction plane
per nucleon and $p_{cm}^{proj}$ is projectile momentum in the center
of mass system (c.m.s.), $y_{cm}$ and $y_{cm}^{proj}$ are the
nucleon rapidity and projectile rapidity in c.m.s., respectively.
The nuclear stopping ratio $R$ as a measurement of degree of
stopping of colliding nuclei at scale impact parameter $b_{0}$,
suggested in Ref. \cite{stopping1,stopping2}, is expressed as
$R(b_{0})=\frac{2}{\pi}{\sum
\nolimits_{i}|p_{ti}|}/{\sum\nolimits_{i}^{}|p_{zi}|} $, where
$p_{ti}$ and $p_{zi}$ are transverse and longitudinal momentum of
the $i$th outgoing particle in the c.m.s., respectively. A colliding
system dependence variable $\rho_{mb}$ is also introduced to be a
normalization factor for later calculations. It is defined as
$\rho_{mb}(A, E_{lab})={MB(0)*u_{cm}^{proj}}/{A^{4/3}} $, where the
$MB(0)$ stands for the meson to baryon ratio for central collision
($b_{0}=0$), $u_{cm}^{proj}=\beta_{cm}^{proj}\gamma_{cm}^{proj}$ is
the spatial component of four-velocity of the projectile in the
c.m.s and $A$ is the mass number of a nuclei in the symmetric
colliding system. The overlapping volume of two collide nuclei and
the nuclear passing time for central collisions respectively satisfy
$V\propto A$, and $t_{pass}={r}/{u_{cm}^{proj}}\propto
{A^{1/3}}/{u_{cm}^{proj}}$, where $r$ is the radius of nuclei. Thus,
we have $\rho_{mb} \propto {MB(0)}/{(V*t_{pass})}$, standing for the
meson to baryon ratio per unit volume, per passing time in the
central collisions, which is used to characterize the strength of
particle production at early stage \cite{MB}.

The UrQMD model \cite{UrQmd} used here is a type of numerical
transport models, which is based on the quark, di-quark, string and
hadronic degrees of freedom. It includes 50 different baryon
species(nucleon, hyperon and their resonances up to 2.11 GeV) and 25
different meson species. Two types of equation of state, the hard
EoS with incompressibility $K=380$MeV(Only for beam energy up to
$4A$ GeV) and cascade are contained in the UrQMD model. The model
has successfully been applied to reproduce the experimental results
from SIS/GSI to SPS/CERN energies \cite{UrQmd2}.

A group of symmetric colliding nuclei with five pairs:
${}^{197}\!$Au+${}^{197}\!$Au, ${}^{129}\!$Xe+${}^{129}\!$Xe,
${}^{96}\!$Ru+${}^{96}\!$Ru, ${}^{58}\!$Ni+${}^{58}\!$Ni,
${}^{40}\!$Ca+${}^{40}\!$Ca are combined with 30 and 12 incident
kinetic energies, respectively. The first combination including
total $150=5\times30$ colliding systems with 30 beam energies per
nucleon from $0.35$ to $200$ GeV (0.35, 0.5, 0.66, 0.83, 1.0, 1.5,
2.0, 3.0, 4.0, 5.3, 6.6, 8.0, 10.0, 10.93, 11.9, 12.9, 13.93, 15,
16.9, 17.92, 18.95, 20, 24.22, 36.0, 55.0, 76.92, 102.33, 131.33,
163.36, 200.0) for each pair of the colliding nuclei and the second
one including total $60=5\times12$ colliding systems with 12 beam
energies per nucleon from $0.35$ to $3.9$ GeV (0.35, 0.5, 0.66,
0.83, 1.0, 1.5, 2.0, 2.35, 2.7, 3.1, 3.5, 3.9) are researched with
cascade and hard EoS of UrQMD, respectively .

\begin{figure}
\centering
\includegraphics[height=18pc,width=22.5pc]{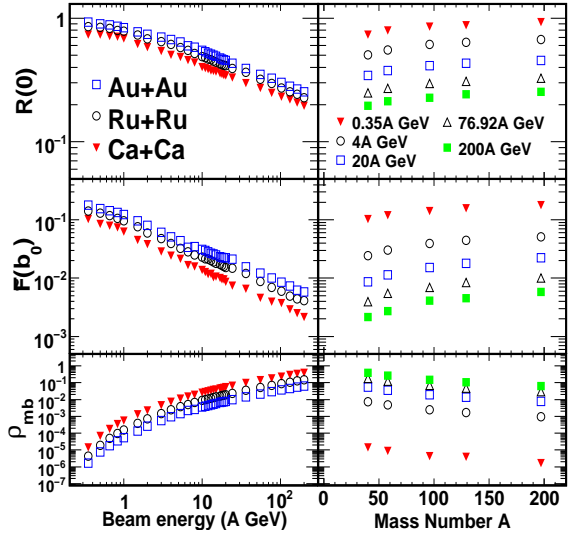}
\caption{(Color Online) Left panels: The excitation function of the
central nuclear stopping ratio, the variable $\rho_{mb}$ and the
semi-central ($0.3<b_{0}<0.4$) scale invariant nucleon sideward
flow. Right panels: System size dependence of the three experimental
observables. } \label{excitation}
\end{figure}

Fig. 1 shows the beam energy and system size dependence of central
($b_{0}=0$) nuclear stopping ratio $R(0)$ and the predefined
variable $\rho_{mb}$ as well as the semi-central ($0.3<b_{0}<0.4$)
scale invariant nucleon sideward flow $\stackrel{\sim}{F}(b_{0})$
with the cascade EoS.  In the left panels of Fig. 1, the $R(0)$ and
$\stackrel{\sim}{F}(b_{0})$ both decrease monotonously with the beam
energy per nucleon from $0.35$ to $200$ GeV for three pairs of
symmetric colliding nuclei: Au+Au, Ru+Ru and Ca+Ca, and larger
nuclear stopping ratio is observed for heavier colliding nuclei than
the lighter one for a fixed beam energy. More detailed information
on the system size dependence of $R(0)$ and
$\stackrel{\sim}{F}(b_{0})$ are illustrated in right panels of Fig.
1. Both $R(0)$ and $\stackrel{\sim}{F}(b_{0})$ increase monotonously
with mass number $A$, which is proportional to the size of the
colliding system. The defined variable $\rho_{mb}$, increasing with
beam energy and decreasing with system size, is also shown in lower
panel of the Fig. 1.

\begin{figure}
\begin{centering}
\includegraphics[height=18pc,width=22pc]{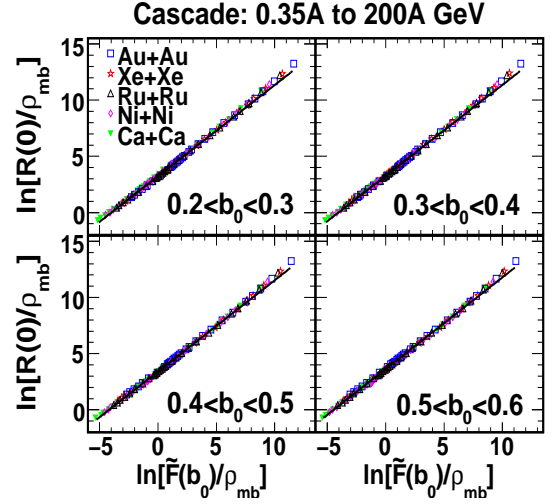}\end{centering}
\caption{(Color Online) The correlation between the central nuclear
stopping ratio and scale invariant nucleon sideward flow, with
$b_{0}$ varying from 0.2 to 0.6 and an interval of 0.1, are
calculated for the mentioned 150 colliding systems within the
cascade EoS. The solid line in each panel is the linear fit of the
corresponding correlation.}\label{cas_corr}
\end{figure}

With cascade EoS in UrQMD, the observables: central ($b_{0}=0$)
nuclear stopping ratio $R(0)$, $\rho_{mb}$ and scale invariant
nucleon sideward flow $\stackrel{\sim}{F}(b_{0})$, with non-zero
$b_{0}$ satisfying $0<b_{0}<0.8$ and with an interval of 0.1, are
calculated for the mentioned 150 colliding systems with beam energy
per nucleon from $0.35$ to $200$ GeV. After the logarithmic
operations are performed on both normalized nuclear stopping
$R(0)/\rho_{mb}$ and normalized scale invariant nucleon sideward
flow $\stackrel{\sim}{F}(b_{0})/\rho_{mb}$, the resulted two
variables, $ln[R(0)/\rho_{mb}]$ and
$ln[\stackrel{\sim}{F}(b_{0})/\rho_{mb}]$ show strong universal
correlation for various colliding systems with a given $b_{0}$ bin.
For illustration, the correlation with the $b_{0}$ from 0.2 to 0.6
and the corresponding linear fit line are shown in Fig. 2. For the
hard EoS case, the results of mentioned 60 colliding systems with
beam energy per nucleon from $0.35$ to $3.9$ GeV are also shown in
Fig. 3. The superposed solid line shown in Fig. 2 and Fig. 3 are the
results of the linear fit for the corresponding correlation.

\begin{figure}
\begin{centering}
\includegraphics[height=18pc,width=22pc]{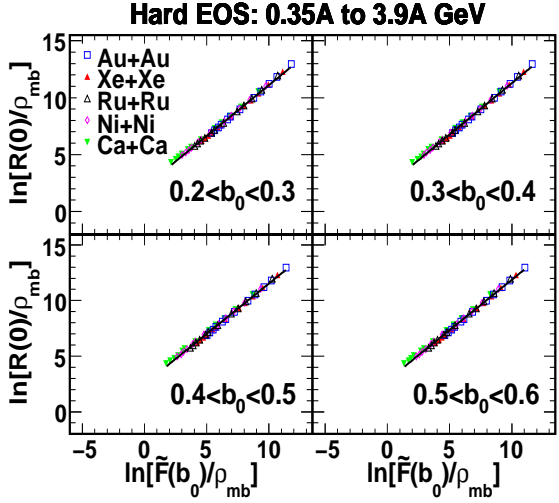}\end{centering}
\caption{(Color Online)  The correlation between the central nuclear
stopping ratio and scale invariant nucleon sideward flow, with
$b_{0}$ varying from 0.2 to 0.6 and an interval of 0.1, are
calculated for the mentioned 60 colliding systems within the hard
EoS. The solid line in each panel is the linear fit of the
corresponding correlation.}\label{EoS}
\end{figure}

By the linear fit of the correlation in Fig. 2 and Fig. 3, the
analytic relation between the two variables: the $R(0)$ and
$\stackrel{\sim}{F}(b_{0})$ can be expressed as:
$$\ln[\frac{R(0)}{\rho_{mb}}]=L\times\ln[\frac{\stackrel{\sim}{F}(b_{0})}{\rho_{mb}}]+m \eqno (1)$$
, where the two fitting parameters $L$ and $m$ are introduced to
represent the slope and intercept, respectively. Generally speaking,
there is nothing particular for any of the two variables $R$ and
$\stackrel{\sim}{F}$, and they are of equal importance. Actually, it
is found that not only central stopping ratio, but also non-central
nuclear stopping ratio is correlated with the scale invariant
nucleon flow $\stackrel{\sim}{F}(b_{0})$, which means for two
independent scale impact parameters $b^{R}_{0}$ and $b^{F}_{0}$, the
relation between the corresponding $R(b^{R}_{0})$ and
$\stackrel{\sim}{F}(b^{F}_{0})$ can be expressed as:
$$\ln[\frac{R(b^{R}_{0})}{\rho_{mb}}]=L\times\ln[\frac{\stackrel{\sim}{F}(b^{F}_{0})}{\rho_{mb}}]+m \eqno (2)$$

\begin{figure}
\begin{centering}
\includegraphics[height=18pc,width=22pc]{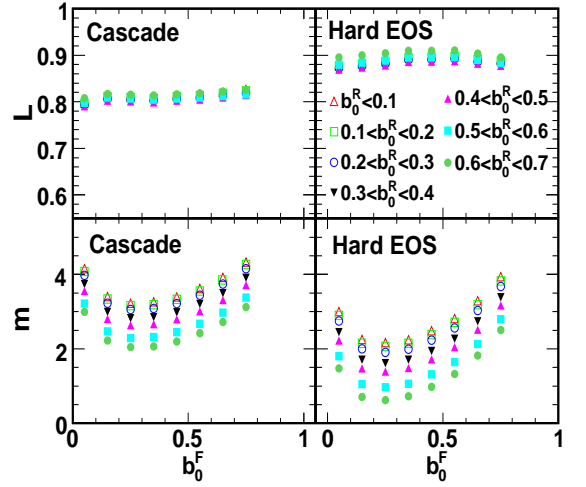}\end{centering}
\caption{(Color Online) The dependence of fitting parameters $L$
(slope: two upper panels) and $m$ (intercept: two lower panels) on
scale impact parameters $b^{F}_{0}$ and $b^{R}_{0}$ for cascade and
hard nuclear EoS.}
\end{figure}

The fitting parameters $L$ and $m$, depending on both $b^{F}_{0}$
and $b^{R}_{0}$ for the cascade and hard nuclear EoS cases, are
illustrated in Fig. 4. In the two upper panels of Fig. 4, the $L$
shows almost no dependence on the $b^{F}_{0}$ and $b^{R}_{0}$ and it
is larger for hard EoS than the cascade one. Thus, it can be
regarded as a constant parameter to characterize the nuclear EoS.
The $b^{F}_{0}$ and $b^{R}_{0}$ dependence of $m$ are also
illustrated in the two lower panels of the Fig. 4, which is strongly
affected by $b^{F}_{0}$, $b^{R}_{0}$ and also by the different
nuclear EoS. The two parameters are both colliding systems
independence, as they are both the universal fitting parameters for
various colliding systems. The parameter $m=m(b^{F}_{0}, b^{R}_{0})$
is only as a function of $b^{F}_{0}$ and $b^{R}_{0}$, and the
so-called correlation function $C(b^{F}_{0},
b^{R}_{0})=e^{-m(b^{F}_{0}, b^{R}_{0})}$ is defined to describe the
correlation strength between the nuclear stopping ratio
$R(b^{R}_{0})$ and scale invariant nucleon sideward flow
$\stackrel{\sim}{F}(b^{F}_{0})$. Two new variables:
$R^{*}(b^{R}_{0})={{R(b^{R}_{0})}/{\rho_{mb}}}$ and
$\stackrel{\sim}{F^{*}}(b^{F}_{0})={\stackrel{\sim}{F}(b^{F}_{0})}/{\rho_{mb}}$
are defined for the simplification of equ. (2). Then, it can be
rewritten as:
$$\stackrel{\sim}{F^{*}}(b^{F}_{0})=\bigg(R^{*}(b^{R}_{0})C(b^{F}_{0},
b^{R}_{0})\bigg)^{\frac{1}{L}} \eqno (3)$$, where the correlation
function, $0<C(b^{F}_{0}, b^{R}_{0})<1$, is only related to the
$b^{F}_{0}$ and $b^{R}_{0}$ for given nuclear EoS.

The colliding system as well as scale impact parameter dependence of
single experimental observable are further investigated for any
specific colliding system. From the equ. (3), for any given
$b^{R}_{0}$ and $b^{F}_{0}$, the two correlative observables are
respectively calculated with two different scale impact parameters,
$(b^{F1}_{0}, b^{F2}_{0})$ and $(b^{R1}_{0}, b^{R2}_{0})$, then we
have:
$$\frac{\stackrel{\sim}{F}(b^{F1}_{0})}{\stackrel{\sim}{F}(b^{F2}_{0})}=\frac{e^{m(b^{F2}_{0},b^{R}_{0})/L}}{e^{m(b^{F1}_{0},b^{R}_{0})/L}}
~;~\frac{R(b^{R1}_{0})}{R(b^{R2}_{0})}=\frac{e^{-m(b^{F}_{0},b^{R2}_{0})}}{e^{-m(b^{F}_{0},b^{R1}_{0})}}\eqno
(4)$$ The terms at the right side of the two equations in (4) are
colliding systems independence and the two equations are satisfied
for any fixed $b^{R}_{0}$ and $b^{F}_{0}$, respectively, which
indicate that the variables of the two observables
$\stackrel{\sim}{F}(b^{F}_{0})$ and $R(b^{R}_{0})$ can be separated
as colliding system dependent term multiplied by scale impact
parameter dependent term as:
\begin{eqnarray}
\setcounter{equation}{5}
R(A, E_{lab}, b^{R}_{0})&=&\xi^{R}(A, E_{lab})\times\eta^{R}(b^{R}_{0}) \\
\stackrel{\sim}{F}(A, E_{lab}, b^{F}_{0})&=&\xi^{F}(A,
E_{lab})\times\eta^{F}(b^{F}_{0})
\end{eqnarray}
With the variable separable property, which is nontrivial and not
common for all the experimental observables, the excitation
properties of the correlative observables for any scale impact
parameter are the same. For better understanding the fitting
parameters $L$ and $m$, as well as the normalization factor
$\rho_{mb}$, the equ. (5) and equ. (6) are introduced into equ. (2),
then we obtain:
\begin{eqnarray}
\ln[\frac{\xi^{R}(A,E_{lab})}{\rho_{mb}}]&=&L\times\ln[\frac{\xi^{F}(A,E_{lab})}{\rho_{mb}}]+m(b^{F}_{0},b^{R}_{0}) \nonumber\\
&&+L\times\ln[\eta^{F}(b^{F}_{0})]-\ln[\eta^{R}(b^{R}_{0})]
\end{eqnarray}
As for various colliding systems ($A, E_{lab}$) and scale impact
parameters $(b^{F}_{0}, b^{R}_{0})$, the equ. (7) are always
satisfied, we have:
\begin{eqnarray}
&&\ln[\frac{\xi^{R}(A,E_{lab})}{\rho_{mb}}]=L\times\ln[\frac{\xi^{F}(A,E_{lab})}{\rho_{mb}}]\\
&&m(b^{F}_{0},b^{R}_{0})=\ln[\eta^{R}(b^{R}_{0})]-L\times\ln[\eta^{F}(b^{F}_{0})]
\end{eqnarray}
The equ.(8) demonstrates that the colliding system dependent terms
of the two correlative observables have been connected by
introducing a proper normalization factor $\rho_{mb}$, which is also
colliding system dependent and may be not unique for present
correlation or even not necessary for other correlative analysis. In
equ. (8), the universal fitting parameter $L$ is uniquely determined
by the colliding system dependent properties of the two observables,
that is the reason the $L$ is not related to the $b^{F}_{0}$ and
$b^{R}_{0}$ for a given nuclear EoS (See Fig. 4). Thus, it is
supposed to be a constant characteristic parameter to characterize
the nuclear intrinsic properties. The variable separable properties
of the two correlative observables in equ. (5) and (6), and the
analytic relation between colliding system dependent terms in equ.
(8) are the origin of the correlation presented by equ. (2). As a
consequence of the correlation for various scale impact parameters,
the intercept parameter $m$ can be expressed as equ. (9), which is
the combination of the scale impact parameter dependent terms of the
two correlative observables and without the cross terms. Derived
from equ. (5), (6) and (9), the differential of the parameter
$m(b^{F}_{0},b^{R}_{0})$ can be written as:
\begin{eqnarray}
\frac{\partial m}{\partial b^{R}_{0}}&=&\frac{\partial
{\ln[\eta^{R}(b^{R}_{0})]}}{\partial b^{R}_{0}}=\frac{\partial \ln[
R(A, E_{lab}, b^{R}_{0})]}{\partial b^{R}_{0}}\\
-\frac{1}{L}\frac{\partial m}{\partial b^{F}_{0}}&=&\frac{\partial
{\ln[\eta^{F}(b^{F}_{0})] }}{\partial b^{F}_{0}}=\frac{\partial
\ln[\stackrel{\sim}{F}(A, E_{lab}, b^{F}_{0})]}{\partial b^{F}_{0}}
\end{eqnarray}
The differential of $m(b^{F}_{0},b^{R}_{0})$ in equ.(10) and (11)
are only related to $b^{R}_{0}$ and $b^{F}_{0}$, respectively (See
Fig. 4) and uniquely determined by the corresponding experimental
observable. The colliding system dependent properties of the
differential of experimental observables in equ. (10) and (11) can
be used to validate whether the observables are variable separable
or not, which is a necessary and not sufficient condition for the
present correlation. Because the equ. (2) is only the fitting
equation for the two correlative observables and not all of the
colliding systems exactly satisfy the equ. (2), the differential of
$m$ may be weakly dependent on specific colliding system.

We have performed correlative analysis between the nuclear stopping
and scale invariant nucleon sideward flow for various colliding
systems and scale impact parameters. The complication of colliding
system dependence of single observable can be separated from the
scale impact parameter dependence, and has been reduced to two
universal fitting parameters $L$ and $m$, which can be used to
determine the nuclear EoS or other intrinsic properties. The
essential of the universal correlation behavior between nuclear
stopping and scale invariant nucleon sideward flow from SIS to SPS
energies may result from the pressure of the matter in HICs, which
is dominated by nuclear EoS, in-medium $NN$ cross section {\it etc.}
and intimately connected to the nuclear stopping and nucleon
sideward flow\cite{MB,NN,meanfield,spsflow,v2,sidewardflow}. The
strong correlation may indicate that the pressure produced in HICs
may be also with the variable separable property as in equ. (8) and
(9), and it may be broken down by the sudden change of the nuclear
bulk properties, such as phase transition {\it etc}. As the phase
transition mechanism is not explicitly involved in UrQMD model and
it is also found in Ref. \cite{collape} that the collapse of
excitation function of the sideward flow and elliptic flow can be
used to probe the first order QGP phase transition. Thus, for
qualitative analysis, it is predicted that the universal correlation
may be broken down at different sets of beam energies for various
colliding nuclei, which could serve as signals for the first order
QGP phase transition. Furthermore, if the universal correlation
pattern restores at much higher energies, where the crossover from
hadronic phase to partonic phase would happen, the location of the
critical point can be unitedly restricted by the mapping of
different colliding nuclei with the corresponding lower limit of the
restoration energies on the QCD phase diagram. The real experimental
data as well as the theoretical calculation are expected to be
compared with the UrQMD simulation results and the detailed
correlation mechanism of the two experimental observables should be
further studied.

\begin{acknowledgements} We thank  Hong-Fang Chen,
Zi-Ping Zhang and Hu-Shan Xu for their helpful discussions. This
work is supported by National Natural Science Foundation of China
(10775131,10675111) and the CAS/SAFEA International Partnership
Program for Creative Research Teams under the grant number of
CXTD-J2005-1.
\end{acknowledgements}


\end{document}